%% file: main.tex
\documentclass[conference,compsoc]{IEEEtran}
\ifCLASSOPTIONcompsoc
  \usepackage[nocompress]{cite}
\else
  \usepackage{cite}
\fi
\ifCLASSINFOpdf
\else
\fi
\usepackage{enumitem}
\usepackage{tikz}
\usepackage{tabularx}
\newcolumntype{Y}{>{\centering\arraybackslash}X}

\newcolumntype{Z}{>{\hsize=1.2\hsize}X}
\newcolumntype{Q}{>{\hsize=.8\hsize}X}
\newcolumntype{V}{>{\hsize=.15\hsize}X}
\usepackage{caption}
\usepackage{subcaption}
\usepackage{stfloats}
\usepackage{multirow}
\usepackage{multicol}
\usepackage{makecell}
\usepackage{balance}
\usepackage{hyperref}
\usepackage{nth}
\usepackage{xurl}
\usepackage{censor}
\usepackage{verbatim}
\usepackage{booktabs}

\begin{document}
%
\title{Children, Parents, and Misinformation on Social Media}


\author{\IEEEauthorblockN{Filipo Sharevski}
\IEEEauthorblockA{School of Computing \\
DePaul University\\
Chicago, IL 60604\\
Email: fsharevs@depaul.edu}
\and
\IEEEauthorblockN{Jennifer Vander Loop}
\IEEEauthorblockA{School of Computing \\
DePaul University\\
Chicago, IL 60604\\
Email: jvande27@depaul.edu}
}


%


\maketitle

\begin{abstract}
Children encounter misinformation on social media in a similar capacity as their parents. Unlike their parents, children are an exceptionally vulnerable population because their cognitive abilities and emotional regulation are still maturing, rendering them more susceptible to misinformation and falsehoods online. Yet, little is known about children's experience with misinformation as well as what their parents think of the misinformation's effect on child development. To answer these questions, we combined a qualitative survey of parents (\textit{n}=87) with semi-structured interviews of both parents and children (\textit{n}=12). We found that children usually encounter \textit{deep fakes}, \textit{memes with political context}, or \textit{celebrity/influencer rumors} on social media. Children revealed they \textit{ask Siri} whether a social media video or post is true or not before they search on Google or ask their parents about it. Parents expressed discontent that their children are impressionable to misinformation, stating that the burden falls on them to help their children \textit{develop critical thinking skills} for navigating falsehoods on social media. Here, the majority of parents felt that schools should also teach these skills as well as media literacy to their children. Misinformation, according to both parents and children affects the family relationships especially with \textit{grandparents} with different political views than theirs.

\end{abstract}


%
\IEEEpeerreviewmaketitle

\input{sections/01.introduction}

\input{sections/02.background}
\input{sections/03.study}
\input{sections/04.results}

\input{sections/05.discussion}
\input{sections/06.conclusion}

\bibliographystyle{IEEEtran}
\bibliography{references}
%

\input{sections/appendix}

\end{document}

%% file: sections/01.introduction.tex
\section{Introduction} \label{sec:introduction}
Social media has become an integral part of the lives of children aged 11 to 17. According to a 2021 census, 32\% of the surveyed children expressed that they couldn't live without YouTube, spending approximately two hours a day on this or other platforms like TikTok \cite{CommonSense2022}. YouTube is the platform that 54\% of teenagers use as their main source of news, in addition to the educational and entertainment content they consume there \cite{CommonSense2019}. But YouTube and TikTok are, at the same time, the platforms that host considerable misinformation \cite{Grant2022, newsguard}, despite their active moderation of falsehoods and inaccurate content \cite{youtube-misinfo, tiktok-safety}.

Though the content appropriateness on social media platforms among children is regulated through age restrictions under the Children's Online Privacy Protection Act (COPPA) \cite{coppa}, it remains relatively easy for children to falsify their date of birth or use older family members' accounts to gain access to social media platforms, potentially exposing them to content that is not suitable for their age \cite{Alomar2022}. In addition, YouTube's and TikTok's algorithmic incentives do appear to distort children’s content in potentially unhealthy ways \cite{paolillo2020youtube, Papadamou2020}, not excluding misinformation \cite{Calder2021, bbc2023}. 

The impact of social media is particularly pronounced in the context of children's development, as they find themselves at a stage where their social and emotional growth leaves them susceptible to influence, and their capacity for self-regulation is still limited \cite{Tartari2015}. Extensive research has thus delved into the various developmental challenges children face as a result of their interactions with (un)suitable content on social media, such as anxiety, depression, dietary concerns, and sleep disturbances \cite{Bozzola2022, Tartari2015, Baldry2019}. For example, the extended use of social media during the early stages of the COVID-19 pandemic caused increased anxiety among both children and their parents \cite{Drouin2020}.

Yet, little to none is known about what misinformation content children encounter on social media and how that affects their development. Furthermore, there is a gap in the understanding of the concerns, experiences, and opinions parents have relative to how misinformation shapes both the interpersonal relationships and the development of their children. To address these gaps, we conducted a two-staged study where we first surveyed 87 parents to answer the following research questions: 

\begin{itemize}
    \itemsep 0.5em
    \item \textbf{RQ1:} What experience do parents have relative to their children's exposure to misinformation on social media? 

    \item \textbf{RQ2a:} What opinions do parents have relative to how children's exposure to misinformation should be handled from a parenting perspective? 

    \item \textbf{RQ2b:} What opinions do parents have relative to how children's exposure to misinformation should be handled from an educational/literacy perspective?

    \item \textbf{RQ3:} How does misinformation affect the development of children, in the opinion of their parents?

\end{itemize}

We followed up with the parents we surveyed to do a semi-structured interview in the second stage of their study, together with their child, in order to broaden our understanding of how parents and children deal with misinformation. Twelve families agreed to do an interview relative to the following research questions: 

\begin{itemize}
    \itemsep 0.5em
    \item \textbf{RQ4:} How \textit{parents} deal with misinformation on social media (encounters, interaction, response)? 

    \item \textbf{RQ5:} How \textit{children} deal with misinformation on social media (encounters, interaction, response)? 

    \item \textbf{RQ6:} How misinformation impacts the relationships within their extended \textit{family}? 

\end{itemize}

We obtained approval from our Institutional Review Board (IRB) to conduct a study with a sample of \textit{n}=87 adult parents who have at least one child in the age group between 11 and 17 years. We focused on this age group in particular because young teens (11-14) and teenagers (15-17) are becoming independent thinkers and are starting to increasingly acquire critical thinking skills \cite{cdc-child-development}. Because children in this age group are a vulnerable population \cite{landrigan2004children}, we carefully and closely followed the recommendation protocols for conducting research with vulnerable populations suggested in \cite{greig2012doing}. First, we obtained formal consent from the parents in the survey, which was anonymous, and participants were allowed to skip any question that felt uncomfortable. The survey took around 20 minutes on average, and we compensated the participants each (total of \$348 for the survey segment). We initially recruited 100 participants for the first segment, but 13 of them did not have a child in the 11-17 age bracket so we had to exclude them from the sample.  

We also obtained consent both from the parents and the children before we conducted the interview where we explained the research objectives and encouraged the children to review the consent forms with their parents before they agreed to be in our study. We used a special consent form in accessible, plain language, so children understand what we were going to ask them, that the study is anonymous, that we don't ask them to access any content, we don't require them to share any details about their time on social media if they feel uncomfortable with both us and their parents, and that they can leave the interview anytime they want without any consequences to them. 

To guarantee the privacy and confidentiality of the parents and children, we offered to do an audio-only interview over Zoom where the parents and children are able to join with an alias (not use their real name(s)). We told them that we would use the audio recording feature only to have a record for transcription. After we got the transcription from the Zoom session, usually no later than 30 minutes from the conclusion of the interview, we checked the transcript for anonymity and immediately deleted the audio recording. Before we started the recording, we ensured our participants that the data we collected in this manner ensured that the privacy risks were minimal and that no personal information was stored. The interviews took around 40 minutes on average, and we compensated each family \$40 for their participation (total of \$480 for the interview segment). 

We communicated with both the parents and children that there are no risks from participation as we are not exposing them to any misinformation content, and we are not asking them to interact with such. We also noted that the parents and children could skip any question they felt uncomfortable answering and that each one of them could take as much as possible time to answer the questions. We offered the participants to remove their answers anytime they wanted. We communicated the findings of our study with each of our participants before we finalized the write-up. None of the parents and children exercised these options and all of them were happy to participate in the research study. We did an extensive debriefing where we talked about the past and current state of misinformation on social media, the perils of exposure to it, and ways to navigate their social media time while encountering misinformation content. We also conducted an extensive debriefing to ensure that none of the parents and children felt that the study incurred any harm to them, that they did not feel uncomfortable during the data collection, and that they felt respected throughout the entire interview. As part of the debriefing, we offered resources where participants could fact-check information they encounter on social media \cite{fact-check, snopes} and we offered both the parents and children that they can contact us about anything they feel they want to know more about related to misinformation anytime after the study.

\noindent \textbf{Findings}. Our study adds important new evidence to the experiences with misinformation of children as a vulnerable population as well as the opinion of their parents relative to their development, particularly: 
\begin{enumerate}
\itemsep 0.3em
    \item Children usually encounter \textit{deep fakes}, \textit{memes with political context}, or \textit{celebrity/influencer rumors} on social media, particularly on YouTube and TikTok; Children revealed they \textit{ask Siri} whether a social media video or post is true or not before they search on Google or ask their parents about it. 

    \item Parents expressed discontent that their children are excessively exposed to misinformation during their time on social media; Parents also voiced a concern that the burden falls \textit{entirely} on them to help their children \textit{develop critical thinking skills} for navigating falsehoods on social media. 
    
    \item The majority of parents felt that schools should also teach these skills as well as media literacy to their children, but some parents were reserved about the misinformation curriculum and the teachers that should offer it to their children.

    \item The majority of parents did feel that misinformation could have \textit{negative consequences} to the development of their children; There were parents that felt misinformation might be \textit{helpful} for children to know early on to avoid memes and fake news.
    
    \item Misinformation, according to both parents and children affects the family relationships especially with \textit{grandparents} with different political views than theirs. 
    
\end{enumerate}

\noindent \textbf{Scope and contributions of this work}: The scope of this work is to gain knowledge, through an unobtrusive approach, about children's and their parents' experiences with misinformation on social media. The contributions, therefore, are fourfold: (1) This is the first work detailing the accounts of \textit{children and their parents} relative to misinformation on social media, particularly YouTube or TikTok as the most popular platforms among children; (2) We provide evidence of the type of content -- deep fakes, memes with political context, and celebrity/influencer rumors -- that children encounter and interact with on social media; (3) Children, our study revealed, are more inclined to ask external resources about potentially fake news than to ask their parents about it; and (4) Parents only have themselves to help their children navigate misinformation both online and in conversations with their extended families, especially with their different-minded grandparents.

%% file: sections/02.background.tex
\section{Background}
\subsection{Children and Misinformation}
Children represent an exceptionally vulnerable population in times when social media permeates everyone's daily lives. Their cognitive abilities and emotional regulation are still maturing, rendering them more susceptible to content that is not always appropriate nor contains factual information \cite{paolillo2020youtube, Papadamou2020, bbc2023}. Children tend to be more emotional, impulsive, and impressionable, making it challenging for them to control their feelings, actions, and time spent on social media \cite{Bhabha2006}. UNICEF's 2021 \textit{Rapid Analysis on Digital Misinformation/disinformation and Children}, for example, highlighted the concern that children worldwide are increasingly exposed to potentially harmful information online, including misleading content shared online \cite{Howard2021}. 

UNICEF's Rapid Analysis report revealed that children are moved by the power of attraction of conspicuous, emotional, or outrageous language used in rumors or manipulations under the guise of reliable information, often by influencers on YouTube and TikTok \cite{neda2021generation}. Moreover, children identify with the influencers they follow and may be at risk of over-trusting and thus sharing mis/disinformation from them \cite{Howard2021}. When it comes to sharing, children want to be popular and that means generating and sharing content geared towards popularity, which often includes rumors around COVID-19, ballot-stuffing in elections, or conspiracy theories around the climate activist Greta Thunberg \cite{dave2020targeting}.

There are few studies concerning children and misinformation, most of which are focused on fake news, misleading information, rumors, and conspiracy theories surrounding the COVID-19 pandemic or health-related issues. Authors in \cite{gotz2020children} surveyed children about COVID-19 rumors and found out that at least 20\% of them believe in alternative explanations about the virus and the appropriate treatments. Another study conducted in Germany in 2020 revealed that 76 percent of 14--24 year-olds have been seeing COVID-19 mis/disinformation at least once a week \cite{Holznagel2017}. A recent study revealed that 41\% of teenagers struggled to identify general health misinformation online, even though this demographic often relies on social media for health-related research \cite{Superlatives2022}. Another study found that secondary school students in half the cases incorrectly identified inaccurate health information without providing any justification for their assessments \cite{Marttunen2021}. Assessing how children aged 11 to 17 together with their parents determine the validity of health information on TikTok, a study found that both parents and children typically distrusted medical information on the platform, even when it came from qualified sources \cite{Feijoo2023}. 

Children's ability to discern the credibility of online information is further challenged by factors such as graphic quality and detailed ``about'' pages, which can sway them into believing that a source is credible. The Stanford History Education Group conducted a study to determine if children were able to determine valid sources of information \cite{Donald2016}. They found that 80\% of the children believed an ad that contained the words ``sponsored content'' was a news source and 30\% of the children believed a fake news account was more trustworthy than an account that was labeled as verified by the social media platform. In a more recent study, the group tested high school students' ability to discern misinformation, showing that two-thirds of them were unable to distinguish news stories from advertisements, over half believed a video of Russian voter fraud was evidence of US voter fraud, and less than 3\% of students evaluated online information through an investigation of sources \cite{doi:Breakstone2021}

\subsection{Parents and Misinformation}
YouTube is one of the most commonly used platforms by both parents and children. While YouTube Kids has been introduced as an effort to control content and protect children from age-inappropriate material, there remains a risk of children encountering unsafe and disturbing content. A recent study found that there was a 3.5\% chance a child would encounter inappropriate content by following YouTube’s recommended videos through ten recommendations \cite{Papadamou2020, Subedar2017}. As children navigate the social media landscape, their responses to and reliance on online information vary with age. Younger children, aged 9-11, often rely on their parents to assist them in determining what is real and fake on social media \cite{Livingstone2014}. In contrast, older children, particularly those over 13, are more likely to handle online encounters independently. Past research also suggests a disconnect between the perceptions of parents and children in these age groups regarding the extent of communication and sharing about online experiences \cite{Wisniewski2017}. Parents, while believing they have a clear understanding of their children's online activities, are often less informed than they assume is the case, leaving their children at a higher risk of exposure to misinformation. 

Parents face the dilemma of how to protect their children from online dangers while simultaneously granting them autonomy. The majority of parents utilize some sort of active and passive monitoring methods to attempt to ensure their children’s safety online \cite{Mai022}. Some parents utilize social media itself to attempt to find ways to monitor their child’s internet use \cite{Wei2022}. A study found that while parents and their older children agreed that there should be some level of privacy for internet use, they frequently disagreed on what the level of privacy should be and what was acceptable for parents to monitor \cite{Cranor2014}. Younger children felt that while their parents implemented rules surrounding internet use, they did not educate them on why the rules were put in place and what content is off limits \cite{Mai022}.

%% file: sections/03.study.tex
\section{Study Methodology}
In circumstances where children are hardly prepared to navigate through misinformative content and their parents are yet to make sense of the parenting approach relative to such content, we set out to investigate both the children's and their parents' perspectives on misinformation on social media. We identified a gap in the general understanding of children's encounters and discernment of misinformation, as well as the experiences with parenting relative to the misinformation's impact on their children's development.

\subsection{Recruitment, Sampling, and Data Collection}
Before we started our recruitment and sampling, we obtained approval from our Institutional Review Board (IRB) to conduct an exploratory survey and follow-up semi-structured interviews (the \hyperref[sec:questionnaire]{survey questionnaire} and the \hyperref[sec:script]{interview script} are provided in the Appendix). Our sampling criteria sought families from the US where both the parents and the children are social media users, the parents are aged 18 or above, and the children were aged between 11 and 17. The criterion ``social media user'' stipulated participants -- both parents and children -- to have at least one social media account and to visit at least one social media platform regularly (daily or weekly) over a period of at least a year, reading at least few posts per visit (there were no restrictions regarding posting, commenting, liking, or reporting posts on platforms as part of ``usage''). As indicated before, we focused on children aged between 11 and 17 because they are a vulnerable population at considerably higher risk of exposure to harmful misinformation \cite{landrigan2004children}. Because children in this age group are becoming independent thinkers and start to increasingly acquire critical thinking skills \cite{cdc-child-development}, we focused on their experiences and the impact misinformation has on their development, without actually exposing them to any harmful content whatsoever. 

We used Prolific for the recruitment, Qualtrics for the survey, and Zoom for the follow-up interviews. We used the Prolific's built-in verification feature to ensure the participants are from the US, adults (i.e., 18 years of age or older), and that they meet the social media user criterion. We included this criterion in order to ensure the validity of our study relative to the familiarity with social media and the equal opportunity for accessing misinformative content as their children (i.e., ensure that parents have the opportunity to access any content their children see online and that they are able to talk about it with their children). We also stipulated that the parents must have at least one child in the age range of 11 -- 17 to which they are lawful parents or legal guardians. We didn't discriminate against other combinations of parenting, children's age, or familiar relationships as long as the above criterion was satisfied (besides, that would have been difficult to ascertain and would have constituted a possible invasion of privacy, which was prohibited by our research protocol approval).

Another stipulation was that that their children are also social media users, i.e., spent some form of supervised time on social media -- either they talk to their parents or share some of the content they see. We focused on the supervision aspect because we wanted to ensure that parents and children have a mutual understanding about the content they see on social media in general and avoid situations where parents, if participating later in the semi-structured interview, face surprises about the content their children interact with online. Equally, we wanted to avoid situations where children are uncomfortable, fearful, or feel they might face any repercussions from their parents because they visited unsupervised content. At the end of the survey, we offered each of the participants -- the parents -- to take part in a flow-up semi-structured interview with their children to speak more about their experiences with misinformation. We allowed them to bring their child that was in the 11 -- 17 age bracket, but they were also free to bring other children to whom they were lawful parents or legal guardians if they wished to (all participants brought only one child). Our protocol also allowed for their other lawful parent or legal guardian(s) to be present in the interview if they wished and we had prepared consent forms for them too (this option was not exercised by our participants). 

The survey was completed by a total of 87 participants, filtering out low-quality responses that didn't answer any of the questions or provided immaterial answers. The filtering was done by each researcher individually and then the research team reached a consensus for the low-quality responses to be removed. Out of these 87 participants who completed the survey, 12 agreed to do the follow-up interview with their children present. We offered them the flexibility to schedule the interview that fits their schedule the best, including weekends, in order to minimize any unnecessary burden to the children (i.e., homework, extracurricular work, etc.). The participants indicated the age of their children as part of the survey, but we also ensured their age as part of obtaining consent before we started the interview to ensure that the children also met the sampling criteria. The survey responses were anonymous, but the interviews initially weren't. 

We offered the participants to use only the audio option in Zoom if they wanted and we recorded the meetings so we would get the automatic transcription for our later data analysis. Once we had the audio transcribed, we removed any personally identifiable information and checked the answers to establish full anonymity. Both the survey and the follow-up interviews allowed the parents and children to skip any question they were uncomfortable answering. The survey took around 20 minutes to complete and the follow-up interviews around 40 minutes. Participants were offered a compensation rate of \$4 for the survey and \$40 for the follow-up interview participation (total for the family). The demographic structure of our survey and interview participants are given in Table \ref{tab:demographics} and Table \ref{tab:interviewdemographics}, respectively. We achieved a balanced and diverse sample, in both the survey and the follow-up interviews. 

\begin{table}[htbp]
\renewcommand{\arraystretch}{1.5}
\footnotesize
\caption{Survey Demographic Distribution}
\label{tab:demographics}
\centering
\begin{tabularx}{\linewidth}{|Y|}
\hline
\footnotesize
 \textbf{Gender} \\\hline
\footnotesize
\vspace{0.2em}
    \hfill \makecell{\textbf{Female} \\ 43 (49\%)} 
    \hfill \makecell{\textbf{Male} \\ 44 (51\%)} \hfill
\vspace{0.2em}
\\\hline

\footnotesize
 \textbf{Parent Age} \\\hline

\footnotesize
\vspace{0.2em}
\hfill \makecell{\textbf{[21-30]}\\ 5 (6\%)} 
    \hfill \makecell{\textbf{[31-40]} \\ 34 (39\%)} 
    \hfill \makecell{\textbf{[41-50]}\\ 37 (43\%)} 
    \hfill \makecell{\textbf{[51-60]} \\ 10 (11\%)} 
    \hfill \makecell{\textbf{[61+]} \\ 1 (1\%)}
\vspace{0.2em}
\\\hline

\footnotesize
 \textbf{Child Age} \\\hline

\scriptsize
\vspace{0.2em}
\hfill \makecell{\textbf{[11]}\\ 11 (11\%)} 
    \hfill \makecell{\textbf{[12]} \\ 23 (22\%)} 
    \hfill \makecell{\textbf{[13]}\\ 15 (14\%)} 
    \hfill \makecell{\textbf{[14]} \\ 15 (14\%)} 
    \hfill \makecell{\textbf{[15]} \\ 13 (13\%)}
    \hfill \makecell{\textbf{[16]} \\ 17 (16\%)}
    \hfill \makecell{\textbf{[17]} \\ 10 (10\%)}
\vspace{0.2em}
\\\hline
\footnotesize
 \textbf{Race/Ethnicity} \\\hline

\footnotesize
\vspace{0.2em}
    \hfill \makecell{\textbf{[Black]} \\ 22 (25\%)} 
    \hfill \makecell{\textbf{[Latinx]}\\ 9 (10\%)}
    \hfill \makecell{\textbf{[White]}\\ 41 (48\%)} 
    \hfill \makecell{\textbf{[More than one ethnicity]} \\ 15 (17\%)} 
\\\hline


\footnotesize
 \textbf{Social Media Sites Used (Parent and Children)} \\\hline

\footnotesize
\vspace{0.2em}
\hfill \makecell{\textbf{Facebook}\\ 60\%} 
    \hfill \makecell{\textbf{Instagram} \\ 55\%} 
    \hfill \makecell{\textbf{YouTube}\\ 49\%} 
    \hfill \makecell{\textbf{TikTok} \\ 41\%} 
    \hfill \makecell{\textbf{X}\\ 28\%} 
    \hfill \makecell{\textbf{Reddit} \\ 22\%} 
    \hfill \makecell{\textbf{Other} \\ 45\%} 
\vspace{0.2em}
\\\hline

\end{tabularx}
\end{table}

\begin{table}[htbp]
\renewcommand{\arraystretch}{1.5}
\footnotesize
\caption{Interview Demographic Distribution}
\label{tab:interviewdemographics}
\centering
\begin{tabularx}{\linewidth}{|Y|}
\hline

\footnotesize
 \textbf{Gender} \\\hline

\footnotesize
\vspace{0.2em}
    \hfill \makecell{\textbf{Female} \\ 5 (42\%)} 
    \hfill \makecell{\textbf{Male} \\ 7 (58\%)} 
    \hfill
\vspace{0.2em}
\\\hline

\footnotesize
 \textbf{Parent Age} \\\hline

\footnotesize
\vspace{0.2em}
    \hfill \makecell{\textbf{[31-40]} \\ 6 (50\%)} 
    \hfill \makecell{\textbf{[41-50]}\\ 5 (42\%)} 
    \hfill \makecell{\textbf{[51-60]} \\ 1 (8\%)} 
\vspace{0.2em}
\\\hline

\footnotesize
 \textbf{Child Age} \\\hline

\scriptsize
\vspace{0.2em}
\hfill \makecell{\textbf{[11]}\\ 7 (59\%)} 
    \hfill \makecell{\textbf{[12]} \\ 1 (8\%)} 
    \hfill \makecell{\textbf{[13]}\\ 1 (8\%)}  
    \hfill \makecell{\textbf{[15]} \\ 2 (17\%)}
    \hfill \makecell{\textbf{[16]} \\ 1 (8\%)}
\vspace{0.2em}
\\\hline
\footnotesize
 \textbf{Race/Ethnicity} \\\hline

\footnotesize
\vspace{0.2em}
    \hfill \makecell{\textbf{[Black]} \\ 1 (8\%)} 
    \hfill \makecell{\textbf{[Latinx]}\\ 1 (8\%)} 
    \hfill \makecell{\textbf{[White]}\\ 9 (76\%)}
    \hfill \makecell{\textbf{[More than one ethnicity]} \\ 1 (8\%)} 
\\\hline

\footnotesize
 \textbf{Social Media Sites Used (Parent and Children)} \\\hline

\footnotesize
\vspace{0.2em}
\hfill \makecell{\textbf{Facebook}\\ 54\%} 
    \hfill \makecell{\textbf{Instagram} \\ 42\%} 
    \hfill \makecell{\textbf{YouTube}\\ 100\%} 
    \hfill \makecell{\textbf{TikTok} \\ 46\%} 
    \hfill \makecell{\textbf{X}\\ 21\%} 
    \hfill \makecell{\textbf{Reddit} \\ 25\%} 
    \hfill \makecell{\textbf{Other} \\ 17\%} 
\vspace{0.2em}
\\\hline

\end{tabularx}
\end{table}

\subsection{Trust and Ethical Considerations}
As this was a study with children as a vulnerable population, it was important for us to establish trust and assurances about the goals of the study and the safeguard protections we had in place. Following the suggestions for considering the ethical aspects when doing research with children as a vulnerable population \cite{greig2012doing}, we communicated that the goal of our study was to capture the ``richness'' of their experiences with any content they encountered on social media that they later realized -- either through checking other information sources or asking their parents -- was false and misinformative. A considerable amount of work is focused on inappropriate content that children see on social media \cite{Papadamou2020, Subedar2017} but we didn't want to focus on this content to avoid any discomfort to the children during any part of the interview. Therefore, we communicated to both the children and their parents that we were not interested in anything disturbing but only in content such as memes, videos, images, and posts, that they later learned were false. 

Here, we offered the option for the children to either ask their parents about such an occurrence they both wanted to speak about or anything they encountered on their own. Prior to doing any of the tasks and conducting the interview, we told the children and their parents that they could ask us to stop the interview, stop the recording, or remove any answers or readings at any point in time. We also shared the interview questions beforehand with them and told the parents and the children that we wouldn't ask any demographics beyond age -- such as the children's gender or ethnicity -- to preserve their anonymity (the parents already disclosed their demographics as part of the survey). Only after we received both parents' and children's explicit permission that they were okay with proceeding with the study and sharing their experiences with misinformation on social media, we began the audio-only recorded Zoom session and proceeded to complete the interviews. We notified each parent and child when we started each audio recording, and we told them that they could take as much time as they needed to answer any questions or share any information they deemed relevant.  

After we collected both children's and parents' answers, we debriefed them about the perils of misinformation on social media and offered them resources where participants could fact-check any content they encounter \cite{fact-check, snopes}. We let the parents and children know that they can contact us anytime after the study if they feel there is anything more about misinformation that they would like to know. To ensure we obtained a correct understanding of their experiences with misinformation, we reviewed the main points we recorded during the interview and clarified any misunderstandings we might have. We also sent a draft of our paper to the parents as the main point of contact for feedback and final approval.

We employed lengthy explanations to ensure our participants that we were not involved with any of the social media platforms such as YouTube or TikTok nor with any of the content they have seen on these platforms (i.e., we never posted, interacted, or asked anyone to do such a thing on our behalf as researchers). We were also careful not to appear in favor nor in support of particular misinformation moderation strategies in order to maintain full researcher impartiality. We communicated that our ultimate goal is to create circumstances where children are protected from dangerous misinformation, where children have the resources, skills, and knowledge to avoid or safely engage with content they believe is false, and that misinformation does not burden their parenting and education. We pointed out that, this goal, however, doesn't prevent from misusing our findings or misinterpreting them in the context of debates for pro/against free speech on social media \cite{gettr-paper}, as misinformation studies have historically been used for such debates and claims have been made that misinformation research community is biased against the conservatives in the United States \cite{Tollefson}. We therefore ensured both the parents and children that we would take all the precautions when writing and reporting the results to avoid this possibility. 

As a safeguard to prevent from any political contextualization of both the results and the survey/interview flow, we deliberately choose not to use political positioning as a sampling criterion, nor to collect or ask any demographic or content-related questions explicitly related to politics. We were aware that misinformation is hard to decouple from interpretation through political or identity lenses \cite{folk-models}, but we worked with children as vulnerable population and we were ethically bound to avoid any unnecessary harm that might come from asking potentially controversial questions. Equally, we communicated with parents that our investigation is only interested with the experiences with and opinions about misinformation outside of any personal preference, values, or political believes.


\subsection{Data Analysis}
We performed an inductive coding approach \cite{Saldana.2013, Thomas.2006} to identify frequent, dominant, or significant aspects both in the answers of our participants in the survey (parents) and the answers of the participants (parent(s) and children) in the interview. One of the researchers open-coded all the data. The resulting codes were then discussed with a second coder, who then independently coded all the data on their own. We reached a reaching a $k = 0.96$ of \textit{Inter-Rater Reliability} (IRR) \cite{Cohen.1960}, which we deemed acceptable. 

We structure a \hyperref[sec:codebook]{codebook}, listed in the Appendix, that captured five main aspects: (i) \textit{encounters and engagement} i.e., codes pertaining to the children's and parents' encounter and engagement with misinformation on social media; (ii) \textit{parenting} i.e., codes related to the parenting approach for children when it comes to misinformation such as regulation and debunking of falsehoods;  (iii) \textit{education} i.e., codes describing the participants' opinions about teaching misinformation as a topic in schools; (iv) \textit{development} i.e., codes describing the participants' opinions about the effect misinformation has on the development of their children; and (v) \textit{misinformation and families}, i.e., codes pertaining to the opinions about how misinformation impacts relationships within the family and with family friends. When reporting the results, we used verbatim quotations to convey the way parents and children social media users experience misinformation. We referenced the quotations in the order participation and role, e.g., \textbf{P24} for participant 24, \textbf{P18-Parent} for parent \textbf{P18}, and \textbf{P18-Child} for their child.

%% file: sections/04.results.tex
\section{Parents: Children and Misinformation}
In the first segment of our study, we conducted a survey in which we asked parents of children between 11 and 17 about their opinions on: (i) their children's exposure to misinformation; (ii) how misinformation affects their parenting; (iii) whether misinformation should be taught in schools; and (iv) how misinformation affects the development of their children. Our goal here was to obtain a broader understanding, from parents' perspective, about how misinformation impacts their children's lives and what mediation strategies are important for parenting and education to ensure proper development in the face of misinformation.

\subsection{RQ1: Children's Exposure to Misinformation}
The first research question aimed to understand the experiences that parents have relative to their children's exposure to misinformation online. The answers from the survey were driven towards two aspects of this exposure: (i) \textit{encounters} of misinformative content; and (ii) \textit{engagement} with such content. The encounters the children had with misinformation on social media were mostly with content that was either \textit{political misinformation} or \textit{rumors about influencers and celebrities}. For example, participant \textbf{P51} pointed out that ``\textit{YouTube is perhaps the most prevalent one for fake news and misinformation as [their] child tends to watch a lot of social and political commentary that skews conservative/alt-right.}'' Participant \textbf{P87} added that their child ``\textit{regularly comes across multiple instances of misinformation in the comments part of YouTube, where individuals frequently exchange unsubstantiated material and engage in rumor-mongering.}'' Relative to the rumors about influencers/celebrities, parents pointed out that their children regularly see ``\textit{fake celebrity deaths}'' (\textbf{P82}) or rumors about ``\textit{controversial figures like Andrew Tate}'' (\textbf{P66}). 

When it came to the engagement with misinformative content on social media, parents reported that their children either \textit{shared} it with their parents/friends or \textit{commented} on a post they believed was misinformation. Participant \textbf{P86}'s child, for example, ``\textit{inadvertently liked or shared content that later turned out to be false, memes, misinformation, or rumors, but we've [the parents] talked to our child about the importance of being cautious online, fact-checking information, and understanding the potential consequences of sharing inaccurate content}.'' Participant \textbf{P72}, relative to commenting, said that their child ``\textit{usually writes under any fake YouTube post comments like `stop lying'},'' while participant \textbf{P48} explained that their child: 

\begin{quote}
\textit{Regularly encounters politicized topics such as LGBTQIA+ issues, abortion, immigration, and politics. As my child doesn't fall far from the tree (myself), they actively comment to accounts who post fake news. They are very adept at spotting misinformation and debunking it.}    
\end{quote}  

We also asked the parents to provide their opinions regarding the very possibility that their children could easily get exposed to misinformation on social media. Almost all the participants expressed \textit{discontent} about the sheer amount of false content that their children see on a daily basis. The discontent about exposure to misinformation, for most of the parents, was because ``\textit{young minds are impressionable}'' (\textbf{P9}), but at the same time children ``\textit{don't yet possess the critical thinking skills to differentiate what is real from fake online}'' (\textbf{P48}). Parents, like participant \textbf{P49}, substantiated their concerns about the dangers of misinformation, stating: 

\begin{quote}
``\textit{I think misinformation is very dangerous for children. Video short sites like TikTok are dangerous. I see young attractive influencers putting forth misinformation or very anti-social retrograde ideas (e.g., anti-trans), and I'm really afraid that young people are going to fall for that manipulation.}''     
\end{quote}

\subsection{RQ2a: Parenting and Misinformation}
The second research question, in part, aimed to understand what opinions parents have about how their children's exposure to misinformation should be handled through parenting. Here, the parents' opinions were driven towards \textit{regulating} the exposure in the first place and \textit{helping} them recognize and avoid falsehoods online. The regulation aspect was seen as mainly as \textit{parental duty} to control what content their children access online because social media platforms ``\textit{as private business will certainly not do that}'' (\textbf{P49}), despite having age restrictions and content control. The use of accounts designed for children on platforms like YouTube or TikTok is of little help, in the opinion of participant \textbf{P57}, because ``\textit{children could either use an adult account or circumvent the restrictions with the help of their friends}.''

In the view of participant \textbf{P27}, ``\textit{it's inevitable that children will see misinformation so it is more important that they know how to interact and deal with it through close supervision by their parents}.'' This could be achieved, per participant \textbf{P30}, ``\textit{by regular interaction with fake news, encouraging the children to read and then verify suspicious content in the presence and with the help of their parents}.'' Participant \textbf{P63} added that they ``\textit{can't keep [their] kids from being exposed to meme culture entirely, so [they] try to make sure that they're only interacting with age-appropriate posts and don't go overboard with offensive humor}.'' Participant \textbf{P29}, here, offered a concrete parenting example: 

\begin{quote}
``\textit{Our children are elementary-aged and I do not feel they are quite at the age where they can determine the truth of what they see online. We do have conversations where, for instance, if I read an article and immediately believed it but then realized it was a parody article, I would show the children and then explain why I believed it and then I researched to realize it wasn't true. This both shows them that I am not perfect and there is no need to be overly embarrassed by being wrong or misled, but also the importance of verifying what you do read or see online. So I wouldn't set my kids loose on the Internet to be exposed to everything out there, but I think allowing them some exposure with parental guidance is good for beginning to build critical thinking skills}.'' 
\end{quote}

We also asked how misinformation affects general parenting, given that parents are also exposed to falsehoods online and ``\textit{could transfer their biases onto their children},'' as participant \textbf{P38} put it. Participant \textbf{P41} said that misinformation ``\textit{has given [them] an opportunity to teach [their] child an important critical thinking skill}'' and helped participant \textbf{P51} ``\textit{to become more aware of fake news so that [they] can keep [their] child from falling victim to false information}.'' Other participants like \textbf{P56} felt that misinformation ``\textit{adds a layer of ridiculous complexity and confusion to conversations}'' with their child and, according to participant \textbf{P18}, ``\textit{brings tensions within the home and the family affairs}.'' Participant \textbf{P57} added that misinformation, even with careful parenting, ``\textit{doesn't ever truly go away from their child's mind because they are keen on repeats something erroneous that [they] read online that was clearly clickbait/fake news.}''

\subsection{RQ2b: Education and Misinformation}
The second research question also aimed to understand what opinions parents have about how their children's exposure to misinformation should be handled through education in schools, in addition to parenting. The majority of the participants strongly felt that children should be taught both \textit{media literacy} and \textit{critical thinking skills} for discerning misinformation online, but lamented the fact that \textit{no such curricula parts exist} in schools across the US. Participant \textbf{P84} believed that ``\textit{school-based education can help children develop critical media literacy skills, enabling them to evaluate online information for credibility, accuracy, and relevance}, which in turn will help children ''\textit{analyze media messages, understand bias and propaganda, and recognize common techniques used to spread misinformation}.'' Participant \textbf{P29} offered an honest account about the need for media literacy to incorporate misinformation as a topic in schools:

\begin{quote}
  ``\textit{I have always thought media literacy is important, but we were wholly unprepared to teach it to our kids, and I was surprised when I realized this isn't really a part of the curriculum in our kid's schools. Therefore when our kids show us memes or talk about a rumor they heard from a YouTuber, we try to guide (not grill) them through the information and why they think it might be funny/true/false. It is an aspect of parenting I didn't really think about before having kids, but I feel it's super important to get these conversations going as organically as possible, and school should help with that}.''
\end{quote}

In equal degree, ``\textit{emphasis should be placed on teaching children critical thinking skills where they learn how to question information, fact-check, and critically assess the credibility of sources},'' according to participant \textbf{P86}. But, participant \textbf{P71} warned that there ``\textit{should be no political bias in such an education, but the curriculum should be more about identifying fake news and how to fact check information yourself}.'' Participant \textbf{P63} thought that ``\textit{kids could benefit from sensitivity training, in a way, so that they recognize when memes are being hurtful and offensive versus simply funny}.'' Structurally, participant \textbf{P49} recommended that ``\textit{the risks of misinformation should probably be taught in a short video format like YouTube Shorts or TikTok, as children seem to engage with that type of presentation the best}.'' It was important, according to participant \textbf{P22}, that the educational material must ``\textit{use age-appropriate, real world examples, and be done with the consent of parents}.''

Few participants were reserved about the misinformation education, feeling that this is ``\textit{something that should be done at home}'' (\textbf{P35}). As for the reasons why, participant \textbf{P32} felt it is not acceptable to have a ``\textit{liberal teacher indoctrinating children with false information and selling it as the truth and vice versa},'' which was bound to happen. Public schools, in the opinion of participants \textbf{P33}, \textbf{P66}, and \textbf{P87} are ``\textit{places that actually teach outright disinformation to children},'' and were reluctant to let ``\textit{teachers impose their opinions about misinformation on their children}'' (\textbf{P44}).

\subsection{RQ3: Development and Misinformation}
In our third research question, we probed the participants about their opinions on how misinformation affects the development of their children. While the majority of the parents were cautious that misinformation could have \textit{negative developmental consequences}, there were participants that believed misinformation has either \textit{no effect} on the way their children develop or \textit{helps them develop critical thinking skills}. Participant \textbf{P60} compared the exposure of misinformation to children to ``\textit{feeding them a diet of junk}.'' In the opinion of participant \textbf{P84}, misinformation affects both the ``\textit{cognitive development as children could maintain a wide range of misconceptions}'' as well as their ''\textit{emotional development as the repeated exposure to sensational or disturbing content cause serious distress}.''

Participant \textbf{P78} was concerned that fake news ``\textit{hurts parents because some children refuse to believe the opposite of what they learned from social media.}'' This tension, in turn, would cause ``\textit{children to become more polarized as young adults},'' in the opinion of participant \textbf{P47}. Participant \textbf{P42} added to this line of reasoning that ``\textit{the false sense of how the world operates gives a rise to hatred, divisions, and even gets people hurt}.'' Similarly, participant \textbf{P33} was concerned that misinformation could cause their children to ``\textit{radicalize, have fear, and believe conspiracy theories}.''

Participant \textbf{P28} felt that misinformation ``\textit{doesn't really affect [their] children, it just makes for some interesting conversations}.'' The lack of negative effect, in the opinion of participant \textbf{P8}, was because ``\textit{their child was aware of memes and fake news so they know not to believe them}.'' Participant \textbf{P38} thought that misinformation ``\textit{would not impact the development of [their] children, unless the fake news pertained to health and nutrition and they were following the bad advice}.'' Other participants, like \textbf{P71}, were of the opinion that misinformation ``\textit{positively affects their development as I believe it helps them learn to think for themselves and be aware of falsehoods}.'' But there were participants like \textbf{P85} that thought misinformation could have both positive and negative impacts on the children's development: 

\begin{quote}
``\textit{On the negative side, the exposure to erroneous or false information, may alter the children's perception of the world, promote distrust in reputable sources, and cause confusion or worry. It can also have an impact on their attitudes and views. On the plus side, exploring and debating such information with the help of parents or educators may help children develop critical thinking skills. The objective is to create an atmosphere in which children learn to be critical thinkers and responsible digital citizens, ensuring that the effect of falsehoods is neutralized by a solid foundation of correct, factual information.}''  
\end{quote}

\section{Families: Interactions with Misinformation}
We offered to do a follow-up interview with our participant and their children to discuss their individual, as well as family interactions with misinformation. Our goal here was to obtain a broader understanding, through first-account testimonies, about how misinformation shapes the relationship between children and their parents and what both children and parents do when encountering misinformation. We also wanted to further expand on parents' opinions about controlling misinformation through education.    

\subsection{RQ4: Parents' Misinformation Interactions}
Our fourth research question sought to understand the interactions \textit{parents} had with misinformation, i.e., encounters with misinformative content, detection of falsehoods, and response to it. Relative to encounters, most of the parents stated they have seen ``\textit{a lot of political misinformation as well as COVID-related conspiracies, rumors, and other falsehoods}'' (\textbf{P10-Parent}). As to how they discern truth from falsehoods about content they encounter online, parents were \textit{privy} to ``\textit{known accounts that spread fake news, memes, misinformation, or rumors}'' to which they either ``\textit{call them out directly on social media or sometimes kept scrolling down if the post is an absurd falsehood},'' as participant \textbf{P79-Parent} noted. Some of the parents, like participant \textbf{P31-Parent}, offered a more detailed account of how they respond to misinformation online: 

\begin{quote}
``\textit{I do my own research. I look for multiple sources for many of the things I see online and use fact-checking sites like Snopes. There is so much misinformation today, it can be very difficult to know if/how accurate things are. Unless I see someone I know and care about spreading things that are very obviously fake, I do not say anything. If I care about the person on a personal level, I respond and say the information they are sharing is inaccurate and give sources that prove its inaccuracy.}''    
\end{quote}

\subsection{RQ5: Children's Misinformation Interactions}
Our fifth research question sought to understand the interactions children had with misinformation, i.e., encounters with misinformative content, detection of falsehoods, and response to it. Relative to encounters, interestingly, the children we interviewed stated they have seen ``\textit{deep fake videos of Mister Beast}'' (\textbf{P28-Child}). One child mentioned they have encountered a ``\textit{deep fake video of drones}'' (\textbf{P7-Child}), and another one mentioned they usually see ``\textit{deep fakes of celebrities on TikTok}'' (\textbf{P79-Child}). Children in our interviews, in equal degree, mentioned they have encountered ``\textit{memes with political context},'' for example ``\textit{about Joe Biden}'' (\textbf{P9-Child}).

As to how they discern truth from falsehoods about content they encounter online, children mostly turned to \textit{online sources} and on rare occasions \textit{asked their parents} about the possibility of a fake post. Several children said they ``\textit{ask Siri if a topic of a post or a video is true or not}'' (\textbf{P7-Child}) or ``\textit{go on Google if something is like big news to search it up just to make sure it's true or not}'' (\textbf{P18-Child}). One of the children, \textit{P31-Child}, during the interview revealed that they ``\textit{sometimes send videos to their parent and ask: hey, do you think this is real stuff like that}.'' Children said they don't usually engage with the deep fakes or memes they encounter, but they mostly ``\textit{share and laugh about them with friends}'' (\textbf{P28-Child}) or sometimes ``\textit{go and read the comments on the posts}'' (\textbf{P9-Child}).

\subsection{RQ6: Families and Misinformation}
Our fifth research question sought to learn more about how misinformation impacts family affairs and the relationships with family friends. The parents, like \textbf{P9-Parent} said they usually talk about content their children encounter or search for videos for their homework: 

\begin{quote}
``\textit{We've probably had a few discussions before. Sometimes when they see stuff on YouTube and they bring it up. We have to have a conversation about whether or not it's right or real. Usually it comes up at the dinner table. They talk about their day and what they've seen and are watching on YouTube. Sometimes it's homework related when they have to find videos for homework, things like that and we have a conversation about what the source was and if it is right. We ask them: Who was the YouTuber that you were listening to? Are they reliable? Do they speak the truth?}''    
\end{quote}

Interestingly, misinformation often came up in conversations with children's grandparents who ``\textit{often have different political views}'' than the parents themselves (\textbf{P79-Parent}). Some of the participants, like \textbf{P8-Parent}, recalled a ``\textit{weird conversation with grandma because she had a different view on immigrants than they do in their family},'' to which the child added that they ``\textit{don't believe in those things about immigrants}'' (\textbf{P8-Parent}). Other participants like \textbf{P7-Parent}, mentioned that misinformation ran deep in the family affairs:

\begin{quote}
``\textit{I have had conversations with extended family members that I used to try to be a little bit more blunt and forceful with those exchanges, because they're family, and I don't have to filter stuff as much. I have since stopped doing that because I would have a better time training grasshoppers to do tricks than I would, convincing my extended family that their conspiracy theories about the Deep State are bonkers insane.}''     
\end{quote}

%% file: sections/05.discussion.tex
\section{Discussion}
Bringing evidence to the growing interest in the impact of misinformation on vulnerable populations, our study revealed important findings that have implications for the children, their parents, educators, and the platforms where children spend considerable time online.

\subsection{Implications: Children}
Using the first-account experiences from the children and the reports from their parents, we add to the evidence that children are exposed to misinformation when spending time on social media \cite{Howard2021, bbc2023, paolillo2020youtube, Papadamou2020}. Children in our study also revealed that they are attracted to influencers and encounter rumors or unreliable information on YouTube or TikTok, as noted in \cite{neda2021generation}. Our interviews with families also revealed, in concordance with the evidence in \cite{dave2020targeting}, that children tend to trust and share content geared towards popularity, often containing nonfactual information \cite{Rinehart2022}. For instance, the children we interviewed indicated they encountered a fake video of MrBeast offering iPhones at discounted prices, and impostor accounts seeking to exploit MrBeast's content to gain followers \cite{Gerken2023}. 

We did not test any actual misinformation claims or content with the children in our study like in \cite{gotz2020children, Donald2016, doi:Breakstone2021, Marttunen2021, Feijoo2023} but we found evidence that children in our study prefer to do search on their own -- by asking Siri or Google - before they talk to their parents about the truthfulness of the deep fakes, memes with political context, and the influencer/celebrity rumors the encounter on social media. Unlike the evidence in \cite{John2022}, we found evidence that children in our study do not evaluate the credibility of online information based on how relatable the source is to their identity or interests. Instead, children in our study shared that they are cognizant that many influencers ``\textit{post just for views and clicks},''  and that ``\textit{crazy things pop up from these accounts and you know just to scroll past and ignore it}'' as \textbf{P28-Child} pointed out.

Overall, our study highlights the need for immediate and adequate resources where children can fact-check and verify content they encounter on social media. There are places, of course, like Snopes or FactCheck.org, but the content there does not always offer facts about influencer rumors or misinformation, less known deep fakes that are popular with kids, or all the memes that are shared on social media platforms. Moreover, the language and formatting of fact-checking is geared toward adults and might not be comprehensible by children. Children do ask Siri or Google, but here again, the recommended content such as verified news sources is most probably inadequate or unappealing to children. We didn't do an in-depth analysis to see if YouTube or TikTok offer misinformation moderation labels to the children's content on their platforms but previous evidence suggests that these labels are ineffective even for adults \cite{Andersen, Ecker, cose2022}. Clearly, some form of revised warnings adapted to children should probably exist too, in order to allow children at least a relatable, factual context to counterbalance the conspicuous, emotional, or outrageous language used in rumors by influences on YouTube or TikTok \cite{neda2021generation, context2022}. Children, past evidence suggests, are impressionable by labels such as ``sponsored content'' on these platforms \cite{Donald2016}, so it is a worthy attempt to additional factual resources to children while they actually see misinformative content.

\subsection{Implications: Parents}
Children in our study, navigating social media, were more inclined to turn to Siri or Google than to their parents to assist them in determining what is real and fake \cite{Livingstone2014}. But they still do talk with their parents about content that might not be factual, our results show. This is an encouraging outcome when it comes to misinformation, given a disconnect exists between the perceptions of parents and children and the extent of communication and sharing about online experiences \cite{Wisniewski2017}. Parents, according to our evidence, appear to have a clear understanding of their children's exposure and interactions with information, though still face the dilemma of how to protect their children from misinformation while simultaneously granting them autonomy. 

As past evidence suggests \cite{Mai022}, the parents in our study revealed they utilize some sort of supervised interaction with fake news, memes, or misleading videos to ensure that their children don't fall victim. Parents in our study reported that they also use their own experiences with misinformation on social media to explain the dangers and ways to discern truth from falsehoods to their children, similarly to other potentially inadequate content \cite{Wei2022}. The balance between privacy for social media use and monitoring didn't explicitly come up in our study \cite{Cranor2014}, but both children and parents expressed that they do have conversations about the rules put in place, what content is off limits, and what content should be always discussed at home \cite{Mai022}.

Parents in our study expressed discontent that their children are exposed to misinformation during their time on social media. While this gave them the opportunity to teach them critical thinking skills, they nonetheless voiced a concern that the burden falls entirely on them, without any safeguards on the platform side nor educational support from schools. Misinformation, in the view of the parents we surveyed and interviewed, does add a layer of complexity to parenting and the effort to ensure children's proper development. Parents, unlike their children, do have fact-checking resources and misinformation moderation labels available to them, but they still need to explain factual information to a level adequate to their children. While resources exist to help parents with ``digital parenting'', offering mediation strategies for potentially harmful content \cite{benedetto2020digital}, no such resources exist specifically for misinformation. 

A need for ``misinformation parenting,' thus, arises from our study. Though in the case of inadequate content, most parents could agree on the criteria and harms, this might not be the case with misinformation. A considerable work shows that people bring their political or other identities to widen the divisions about misinformation engagement and regulation online \cite{Linden2020you} and this runs the risk of disagreements on how to approach misinformation for children to begin with. Even if this is solved and agreed upon, misinformation rapidly and unpredictably arises on social media based on emerging controversial topics or events \cite{Singhal}, making it difficult to maintain relevant resources and mediation strategies that are topic, event, or platform agnostic. On top of all this, misinformation -- with its sheer breadth and multiple modalities -- necessitates topic or event-specific adjustments to parenting strategies according to the child's age and parent’s media competence and beliefs \cite{HAMMER2021106642}.

\subsection{Implications: Education}
Perhaps not entirely surprising, teaching children about misinformation in schools was the topic that parents were divided about. As we already indicated, misinformation in the US is a highly divisive topic, and its impact on children is no exception to it. While our study revealed that the majority of parents do welcome educational interventions about misinformation such as teaching media literacy and critical thinking skills for discerning falsehoods, some parents were skeptical that it would be difficult to decouple these interventions from the current political context. Parents deem it important for their kids to learn media literacy and critical thinking skills but also understand the techniques for spreading misinformation, and the way misinformation is generated around controversial topics or events. At the same time, parents were concerned that teachers would bring their political bias into a misinformation curriculum and that public schools ``\textit{already teach out-right disinformation to children}'' per participants \textbf{P33}, \textbf{P66}, and \textbf{P87}. 

A new law enacted in California shows that misinformation is a worthy topic for child education. The law requires K-12 schools to add media literacy to the curriculum for English language arts, science, math, and history-social studies, including lessons for recognizing fake news and thinking critically about what children encounter on the internet \cite{assembly-bill-873}. This bill doesn't specify the actual lessons as educational interventions, which, according to our findings should be tailored to avoid politicization or create the impression of any form of bias. It is indeed easy to show political misinformation, but our study reveals that children do see deep fakes, memes, and rumors about influencers so our suggestion is to use this particular content and examples from social media (the children in our study pointed out to the deep fakes of the popular influencer Mr. Beast \cite{Gerken2023}) for recognizing fake news. Children in our study did point out they see memes with political content, and we deem it appropriate to use these examples as content that children should altogether avoid or ignore when on social media. The disengagement with political memes is not to preclude critical thinking -- which should be taught in the context of why memes go viral, where they come from, and who benefits from them -- but more so to avoid a risk of exposure to potentially harmful and toxic memes that are found in abundance on social media platforms \cite{beran2019came}. 

Towards the recognition of fake news, the lessons could also adopt a game-based approach, popular with children for learning media literacy and cybersecurity \cite{Zhang-Kennedy}. Such games for recognition of fake news already exist \cite{Grace, Roozenbeek}, but they are predominately based on political misinformation and therefore largely inadequate for children. This is an open area for innovation, considering that the children in our study turned to Siri and Google to recognize fake news. To this point, such games could incorporate interactive elements with voice assistants in addition to, or replacing, the multiple-choice approach of the current game when participants guess whether a social media post is fake news or not. Towards critical thinking about misinformation, the mediation approaches shared in our study by the parents could also be of interest to curriculum designers. For example, it is well known that stories help in raising phishing and cybersecurity awareness \cite{Wash2018} so this could be used to deliver the stories that parents shared in our study (e.g., encountering parody or misleading articles as well as difference-of-opinion with grandparents about immigration rumors and falsehoods) or other experiences with misinformation.

\subsection{Implications: Platforms}
Confirming the past evidence that a child would encounter inappropriate content by following YouTube’s recommended videos \cite{Papadamou2020, Subedar2017}, our study revealed that children do regularly encounter misinformation on social media. We understand that the algorithmic recommendation, paired with aggressive content moderation, even for places like YouTube Kids is not 100\% proof to eliminate deep fakes, rumors, or memetic content. In fact, YouTube themselves confirmed that they could not provide a ``100\% protection for children from age-inappropriate material'' \cite{Subedar2017}. But that, in our opinion, is not a reason not to attempt and provide other platform affordances for children to be able to avoid or safely engage with potentially misleading content (same applies for TikTok or other platforms that children use \cite{Calder2021}).

Platforms, when it comes to adults, balance between removing content, barring the visibility of accounts/content to the others on the platform (i.e., ``shadowbanning''), and applying warning labels that link to authoritative medical, electoral, and community-checked facts \cite{youtube-misinfo}. The reception of this moderation effort has been mixed and platforms have been accused of censorship, restraint on the users' voice, political bias, and the suppression of free speech \cite{folk-models}. This sentiment could well be maintained by the parents -- as our study revealed disagreements about interventions for misinformation -- though we believe that children could at least benefit from the use of warning labels, if not the other forms of moderation. 

Past research suggests that warning labels with an explanation stating why a post is misinformation do help users correctly discern the truthfulness of a topic \cite{context2022}. Such context, written in age-appropriate language could be of benefit to children. For example, warning labels could be applied to deep fakes stating: ``\textit{videos like this are created using Artificial Intelligence algorithms to make it appear as if someone is doing or saying something they didn't. If you feel uncomfortable with it, please talk to your parents, a trusted adult, or read more about it on \hyperlink{https://en.wikipedia.org/wiki/Deepfake}{Wikipedia}.}'' (assuming these videos are detected and handled as such by the platforms accoding to their policies in the first place \cite{Pu2023})

Platforms are also known to provide notable benefits for children, especially in educational settings, where they use them to conduct research, collaborate on homework assignments, and exchange ideas, enhancing their learning experience. Towards the goal of parenting and educational support about misinformation, parents recommended that ``\textit{the risks of misinformation should probably be taught in a short video format like YouTube Shorts or TikTok, as children seem to engage with that type of presentation the best}'' according to participant \textbf{P49}. We welcome this idea and believe that the recommended educational interventions or parental mediation approaches could also benefit from such short videos where examples of recognizing fake news, discerning deep fakes and memes, or dispelling a rumor are recommended to children when searching content around controversial or age-inappropriate topics. Using the example where medical professionals take it on themselves to dispel health-related misinformation on TikTok \cite{Raphael}, for example, we believe that teachers or even the influencers popular with children could be invited to create such videos.

\subsection{Limitations}
Our research was limited in its scope to families from the US where both the parents and the children are social media users, the parents are 18 or above age, and the children were aged between 11 and 17. Another limitation is that we sampled English-speaking families and the results might not apply to families that speak other languages or see content in other languages on social media. It is especially important to acknowledge this limitation as recommendation algorithms, influencers, or other ways of content engagement on social media differ based on language and personalization factors \cite{Boeker}, and as such, might yield our findings entirely irrelevant. Though we had a balanced and representative sample, we acknowledge that other and larger samples could yield results and findings that differ from ours. Similarly, a limitation comes from the recruitment platform and the survey provider. A limitation comes from the survey and the interview instruments and other qualitative studies that might yield variations in the experiences with misinformation, differences in the opinions about parenting mediation strategies or educational interventions, or misinformation effects on children's development.

We didn't test any recognition of fake news, literacy, or critical thinking skills with actual social media misinformation, with either the parents or the children. This limits the results of our study from generalization beyond what was present as misinformation on social media platforms during the time we conducted the study -- the second half of 2023. We implicitly ensured with both the parents and the children about what content constitutes misinformation, but there is a possible limitation to the results where an experience is shared with content that might not have been false, misleading, or inaccurate. We had no option to ensure this was not the case as we did not ask the parents and children to share any content. 

Our study is also limited to the state of the social media platforms that both children and parents use. Future platforms, or changes in the misinformation moderation policies of platforms, might affect how parents and children encounter, engage, and interact with misinformation. We also didn't ask parents nor suggest using any alternative applications that offer safer access to social media as we wanted to gather the natural experiences with misinformation as-is on social media. The use of such applications, restrictions, or other types of parental filtering might well affect the content children encounter on social media and, as such, produce variations in the findings.

Though we left both the parents and the children sufficient time to express their opinions, answer their questions, and comment on aspects they deemed important, nonetheless, the time allowed might have not been sufficient for them to formulate a more informed expression about their overall experience with misinformation on social media. Especially for the children, there is a limitation to the formulation and the choice of answers because we stipulated that they talk about supervised encounters with misinformation. There might have been instances where they might have encountered misinformation without the supervision of their parents and our findings do not pertain to these experiences. 

For the parents, a limitation comes from our choice to avoid using any politically related questions or sampling criteria. Misinformation is hard to decouple from a political context, but our findings could not be generalized relative to the political dichotomy in the US. Despite all these limitations, and similar to mixed and qualitative studies in general, we believe our results provide rich accounts of both children's and their parents' lived experiences with misinformation on an unprecedented scale, research design, and academic comprehensiveness.

\subsection{Future Work}
As this is among the first studies investigating the children's and their parents' experiences with misinformation, we are interested in pursuing several research directions stemming from the findings. We would like to replicate our study with a larger sample and possibly test some misinformation content to learn more about ways children discern truth from falsehoods online, especially for deep fakes, memes with political context, and rumors about influencers/celebrities. We are also interested in working with parents in developing ``misinformation parenting'' resources and testing their reception and effectiveness in mediating misinformation exposure to their children. In equal degree, we would work to create and test misinformation lessons, short videos, and educational interventions such as stories with children in classrooms. The misinformation moderation on social media, as an active research area, is also of our interest and we are planning to develop and test various warning labels in combination with content, specifically designed for children. We welcome any collaboration and suggestions in any of these research directions from the broader misinformation and cybersecurity community as we deem the protection of children is a worthy goal for pursuit.



%% file: sections/06.conclusion.tex
\section{Conclusion}
In this paper, we worked with 87 parents and 12 children to gather their experiences with misinformation on social media as well as learn their opinions about the impact of misinformation on children's development. Our study reveals that children encounter deep fakes, memes with political context, and celebrity/influencer rumors. We learned that children prefer asking Siri or Google before talking to their parents about whether content is misinformation or not. Parents shared that they see misinformation both as a danger and as an opportunity to teach children critical thinking skills, but there wasn't a consensus about whether misinformation should be formally taught in schools. Based on our findings, we proposed courses of action relative to the children's time on social media, parenting mediation strategies, and educational interventions. We believe our qualitative results illuminate a compelling set of actionable items, and as such, provide the locus for further inquiries about the cognitive and emotional development, safety, and decision-making of children in presence of misinformation on social media.

%% file: sections/appendix.tex
\section*{Appendix}
\subsection*{Study Questionnaire} \label{sec:questionnaire}

\begin{enumerate}
\itemsep 0.5em

        
        
        \item Has your child(ren) \textbf{encountered} any content on these sites they said was \textbf{fake news, memes, misinformation,} or \textbf{rumors}? Please elaborate on your response. Avoid answering only with “yes” and “no”. 
        
        \item Has your child(ren) \textbf{posted} or \textbf{interacted} with \textbf{fake news, memes, misinformation,} or \textbf{rumors} on any social media sites that you know of? Please elaborate on your response. Avoid answering only with “yes” and “no”. 

        \item What is your opinion regarding child(ren)'s \textbf{exposure} to \textbf{fake news, memes, misinformation,} or \textbf{rumors} on social media sites? Please elaborate on your response. 
        
        \item Have you \textbf{talked} with your child(ren) about \textbf{fake news, memes, misinformation,} or \textbf{rumors}? If yes, in what context? If not, why not? 

        

        \item In your opinion, should child(ren) be \textbf{protected from fake news, memes, misinformation,} or \textbf{rumors} on social media sites and \textbf{how}? Please elaborate on your response. Avoid answering only with “yes” and “no”.

        \item How does \textbf{fake news, memes, misinformation,} or \textbf{rumors}, in your opinion, \textbf{affect the parenting} of your child(ren)? Please elaborate on your response.

        \item In your opinion, should child(ren) be \textbf{educated in school} about the \textbf{risks of fake news, memes, misinformation,} or \textbf{rumors} on social media sites and \textbf{how}? Please elaborate on your response. Avoid answering only with “yes” and “no”.

        \item How does fake \textbf{news, memes, misinformation,} or \textbf{rumors}, in your opinion, \textbf{affect the development} of your child(ren)? Please elaborate on your response.

\end{enumerate}

\subsection*{Interview Script} \label{sec:script}
\begin{enumerate}
\itemsep 0.5em

\item What social media sites or platforms do you usually use the most? State all that apply (each of the family members).




\item Have you \textbf{encountered} any content on social media platforms that was fake news, memes, misinformation, or rumors (each of the family members)?

\item Have each of you \textbf{posted} or interacted with fake news, memes, misinformation, or rumors on any social media sites that you know of?

\item Have you talked \textbf{within your family} about fake news, memes, misinformation, or rumors? If yes, in what context? If not, why not?

\item Have you had any \textbf{agreements or disagreements} about fake news, memes, misinformation, or rumors? If yes, in what context? If not, why not?

\item How each of you yourself \textbf{recognize and respond} to fake news, memes, misinformation, or rumors on social media sites? 

\item How does fake news, memes, misinformation, or rumors, in your opinion, \textbf{affect the affairs in your family}?

\item How does fake news, memes, misinformation, or rumors, in your opinion, \textbf{affect the affairs with your friends}?


\item Would you like to add anything else we didn't asked about fake news, memes, misinformation, or rumors?

\item Demographic questions. What is your: age, gender, ethnicity/race, education, and computer proficiency. 

\end{enumerate}

\subsection*{Codebook} \label{sec:codebook}

\begin{itemize}
    \item \textbf{Encounters and engagement} Codes pertaining to the child(ren)'s and parent's encounter and engagement with misinformation on social media.
    
    \begin{itemize}
        \item \textbf{Encounters} Codes pertaining to the \textit{encounters} of misinformation on social media.
        
        \begin{itemize}
            \item \textbf{Memes} The participant expressed that they encountered memes with political context on social media 
            \item \textbf{Rumors about celebrities or influencers} The participant expressed that they encountered rumors about celebrities or influencers on social media
            \item \textbf{Deep Fakes} The participant expressed that they encountered synthetic video and/or audio on social media
           \item \textbf{Conspiracies} The participant expressed that they encountered conspiracies on social media
        \end{itemize}

        \item \textbf{Engagement} Codes pertaining to the \textit{engagement} and interaction with misinformation on social media.
        
     \begin{itemize}
            \item \textbf{Shared a post} The participant expressed that they a misinformation post from social media 
            \item \textbf{Commented on a post} The participant expressed that they commented on a misinformation post on social media
            \item \textbf{Ask Siri} The participant expressed that they asked Siri about the truthfulness of a post on social media
            \item \textbf{Ask Google} The participant expressed that they searched Google to find out about the truthfulness of a post on social media
            \item \textbf{Ask a Parent} The participant expressed that they asked their parent about the truthfulness of a post on social media
        \end{itemize}

    \end{itemize}
    
    \item \textbf{Parenting} Codes pertaining to the parenting approach for children when it comes to misinformation such as regulation and debunking of falsehoods.
        
        \begin{itemize}
            \item \textbf{Exposure regulation} The participant expressed that they are on the opinion that children's exposure on social media should be tightly regulated
            \item \textbf{Help recognize and avoid falsehoods} The participant expressed that they are on the opinion that parents should help children recognize and avoid falsehoods on social media
        \end{itemize}

    \item \textbf{Education} Codes pertaining to the educational approach for children when it comes to teaching misinformation in schools.
        
        \begin{itemize}
            \item \textbf{Media Literacy Skills} The participant expressed that they are on the opinion that children should be taught media literacy skills at school
        \item \textbf{Critical Thinking Skills} The participant expressed that they are on the opinion that children should be taught critical thinking skills at school
        \item \textbf{Misinformation should be taught at schools} The participant expressed that they are on the opinion that children should not be taught anything related to misinformation at school

        \end{itemize}

    \item \textbf{Development} Codes pertaining to the participants' opinions about the misinformation has on the development of their child(ren).

        \begin{itemize}
            \item \textbf{No effect} The participant expressed that they are on the opinion that misinformation on social media has no effect on their children's development 
        \item \textbf{Teaches them critical thinking skill} The participant expressed that they are on the opinion that misinformation on social media helps their children develop critical thinking skills 
        \item \textbf{Negative effect} The participant expressed that they are on the opinion that misinformation on social media has negative effect on their children's development 
        \end{itemize}

    \item \textbf{Misinformation and families} Codes pertaining to the participants' opinions about the misinformation's impact on the relationships within the family and with family friends.).

        \begin{itemize}
            \item \textbf{Conversations with children} The participant shared that misinformation on social media is a topic discussed at home with their children
        \item \textbf{Conversations with grandparents} The participant shared that misinformation on social media is a topic discussed with the children's grandparents
        \end{itemize}

\end{itemize}